\documentclass[12pt,reqno]{amsart}%
\usepackage{amssymb,amsbsy}
\usepackage{amsfonts,amsbsy}
\usepackage{amssymb}
\usepackage{graphicx}
\usepackage{amsfonts}
\usepackage{amsmath}%
\usepackage{url}%
\providecommand{\U}[1]{\protect\rule{.1in}{.1in}}
\newcommand{\be}{\begin{equation}}
\newcommand{\ee}{\end{equation}}
\newcommand{\ba}{\begin{array}}
\newcommand{\ea}{\end{array}}

\newtheorem{prop}{Proposition}
\newtheorem{cor}{Corollary}
\begin{document}
\title[Reciprocal transformations for St\"ackel-related Liouville integrable systems]{Reciprocal
transformations\\for St\"ackel-related Liouville integrable systems}
\author{Maciej B\l aszak}
\author{Artur Sergyeyev}
\address{Institute of Physics, A. Mickiewicz University\\
Umultowska 85, 61-614 Pozna\'{n}, Poland}
\email{blaszakm@amu.edu.pl}
\address{Silesian University in Opava, Mathematical Institute, Na
Rybn\'\i {}\v{c}ku 1, 746\,01 Opava, Czech Republic}
\email{Artur.Sergyeyev@math.slu.cz}  \maketitle

\begin{abstract}
We consider the St\"ackel transform, also known as the
coupling-constant metamorphosis, which under certain conditions
turns a Hamiltonian dynamical system into another such system and
preserves the Liouville integrability. We show that the
corresponding transformation for the equations of motion is nothing
but the reciprocal transformation of a special form and we
investigate the properties of this transformation. This result is
further applied for the study of the $k$-hole deformations of the
Benenti systems or more general seed systems.


\end{abstract}





\section*{Introduction}

The St\"ackel transform \cite{bkm}, also known as the
coupling-constant metamorphosis \cite{hiet} (cf.\ e.g.\ also \cite{ts}),
is a powerful tool for producing new
Liouville integrable systems from the known ones. This is
essentially a transformation that maps an $n$-tuple of functions in
involution on a $2n$-dimensional Poisson manifold into another
$n$-tuple of functions on the same manifold, and these $n$ new
functions are again in involution. In the present paper we
show that the corresponding transformations for
equations of motion are nothing but reciprocal transformations. We also study
the properties and present some applications of the latter.

The significance of reciprocal transformations in the theory of integrable
nonlinear partial differential equations is well recognized. These
transformations were intensively used in the theory of dispersionless systems
as well as the theory of soliton systems (see e.g.\ \cite{cr2,cr1} and
references therein). \looseness=-1
However, the role of the reciprocal transformations in the theory of
finite-dimensional dynamical systems is far from being fully explored, and the goal
of the present paper is to contribute to such an exploration by developing the
theory of reciprocal transformations for Liouville integrable Hamiltonian
systems. To the best of our knowledge, such transformations first appeared in
the paper \cite{hiet} by Hietarinta et al., where the concept of the
coupling-constant metamorphosis, or the St\"ackel transform \cite{bkm}
(cf.\ also \cite{ves} for even more general transformations in the action-angle variables and
\cite{kkmw, kkmw2,kkmw3,ts} for more recent developments), was introduced.
The reciprocal transformation appeared in this context as a transformation
expressing the time (evolution parameter) for the target system through that
of the source system \cite{hiet}, but the question of whether it sends the
(solutions of) equations of motion for the source system into those of the
target system was not addressed in \cite{hiet}. \looseness=-1

In fact, as we show below, this transformation, when applied to the equations
of motion of the source system, in general does \emph{not} yield the equations
of motion for the target system, unless we restrict the equations of motion
onto the level surfaces of the corresponding Hamiltonians, see Propositions
\ref{eomp} and \ref{eomp1} below for details.

Even more broadly, we show that two Liouville integrable systems related by an
appropriate St\"ackel transform for the constants of motion are
related by the reciprocal transformation for the equations of motion
restricted to appropriate Lagrangian submanifolds (see e.g.\ Ch.3 of
\cite{cas} and references therein
for more details on the latter).

Moreover, we present a multitime extension of the original reciprocal
transformation from Hietarinta et al.\ \cite{hiet}, and study the applications
of this extended transformation to the integration of equations of motion in
the Hamilton--Jacobi formalism using the separation of variables
(cf.\ \cite{bkm}).

As a byproduct, we present reciprocal transformations for a large class of
dispersionless, weakly nonlinear hydrodynamic-type systems, the so-called
Killing systems \cite{blasak} that are intimately related to the
St\"ackel-separable systems \cite{f1,f2,m}.




In the second part of the paper we consider the relations among
classical Liouville integrable St\"ackel systems on
$2n$-dimensional phase space. In \cite{mac2005} infinitely many
classes of the St\"ackel systems related to the so-called seed
class,
namely, the $k$-hole deformations of the latter, were constructed. Here we show
that any $k$-hole deformation consists of a sequence of elementary
deformations (one-hole deformations). These elementary deformations are
nothing but particular cases of the St\"ackel transforms considered in the
first section of the present paper.
Hence the equations of motion for infinitely many classes of St\"ackel systems
are related, upon restriction onto appropriate Lagrangian submanifolds, to the
equations of motion for the systems from the seed class (which is a natural generalization
of the Benenti class for the classical St\"ackel systems)
by a sequence of reciprocal transformations.
\looseness=-1

The significance of this result stems from the fact that the overwhelming majority of
known today Liouville integrable natural dynamical systems that admit
orthogonal separation of variables belong to the seed class. The $k$-hole
deformations of such systems again are Liouville integrable natural dynamical
systems and admit orthogonal separation of variables, but the corresponding
separation curves no longer are of the seed type. Our results make it possible
to understand how the corresponding dynamics is different from that of the
systems from the seed class, and thereby reveal new properties of the deformed
systems, which will be discussed in more detail elsewhere.


\section{Main results}


We start with the following simple results that slightly generalize
Proposition 2 from \cite{ts} and the results of \cite{hiet, kkmw}. The proof
is by straightforward computation.

\begin{prop}
\label{trp} Let $(M,P)$ be a Poisson manifold with the Poisson bracket
$\{f,g\}=(df, Pdg)$. Consider $k$ functions on $M$ of the form
\begin{equation}
\label{df}H_{i}=H_{i}^{(0)} +\alpha H_{i}^{(1)},\quad i=1,\dots,k,
\end{equation}
where $\alpha$ is a parameter, and $H_{i}^{(0)}$ and $H_{i}^{(1)}$ are smooth
functions on $M$. Assume that $H_{i}$ are functionally independent for all
values of $\alpha$.

Suppose that there exists an $s\in\{1,\dots,k\}$ such that $H_{s}^{(1)}\neq0$
and
\begin{equation}
\label{com0}\{H_{s}, H_{j}\}=0
\end{equation}
for all $j=1,\dots,k$ and for all values of $\alpha$, and let
\begin{equation}
\label{tr}%
\begin{array}
[c]{l}%
\tilde H_{i}=\tilde H_{i}^{(0)}+\tilde{\alpha}\tilde H_{i}^{(1)}=H_{i}^{(0)} -
H_{i}^{(1)} (H_{s}^{(0)}-\tilde\alpha)/H_{s}^{(1)}, \quad i=1,\dots
,s-1,s+1,\dots,k,\\[3mm]%
\tilde H_{s}=\tilde H_{s}^{(0)}+\tilde{\alpha}\tilde H_{s}^{(1)}=-(H_{s}%
^{(0)}-\tilde\alpha)/H_{s}^{(1)},
\end{array}
\end{equation}
where $\tilde\alpha$ is another parameter.

Then we have
\begin{equation}
\label{com0a}\{\tilde H_{s}, \tilde H_{j}\}=0
\end{equation}
for all $j=1,\dots,k$ and for all values of $\tilde\alpha$.
\end{prop}

\begin{cor}
\label{com} Under the assumptions of Proposition 1 suppose that we have
$\{H_{i}, H_{j}\}=0$ for some (fixed) $i$ and $j$ and for all values of
$\alpha$.
Then
\[
\{\tilde H_{i}, \tilde H_{j}\}=0
\]
for all values of $\tilde\alpha$.
\end{cor}


The transformation from $H_{i}$ to $\tilde H_{i}$ is known as a
\emph{coupling-constant metamorphosis} \cite{hiet} or
as a (generalized) \emph{St\"ackel transform} \cite{bkm, kkmw}.

>From Proposition~\ref{trp} and Corollary \ref{com} it is immediate that the
transformation (\ref{tr}) preserves (super)integrability: if the dynamical
system associated with $H_{s}$ is Liouville integrable (so $k\geq n\equiv
\frac12\mathop{\rm rank}P$ and $H_{s}$ belongs to a family of $n$ commuting
Hamiltonians $H_{i}$ such that $Pd H_{i}\neq0$ for all $i$) or superintegrable
(i.e., $H_{s}$ is Liouville integrable and $k>n$), then so is the dynamical
system associated with $\tilde H_{s}$. \looseness=-1


Note that the relations (\ref{tr}) can be inverted:
\begin{equation}
\label{itr}%
\begin{array}
[c]{l}%
H_{i}^{(0)}=\tilde H_{i}- H_{i}^{(1)} \tilde H_{s}, \quad i=1,\dots
,s-1,s+1,\dots,k,\\[3mm]%
H_{s}^{(0)}=-H_{s}^{(1)}\tilde H_{s} +\tilde\alpha.
\end{array}
\end{equation}

Recall that the equations of motion associated with a Hamiltonian $H$ and a
Poisson structure $P$ on $M$ read (see e.g.\ \cite{mb})
\begin{equation}
\label{eom0}d x^{b}/dt_{H} =(X_{H})^{b},\quad b=1,\dots,\dim M,
\end{equation}
where $x^{b}$ are local coordinates on $M$, $X_{H}=P dH$ is the Hamiltonian
vector field associated with $H$, and $t_{H}$ is the corresponding evolution
parameter. \looseness=-1

Consider the equations of motion (\ref{eom0}) for $H=H_{s}$ and $t_{H}=t$ and
for $H=\tilde H_{s}$ and $t_{H}=\tilde t$:
\begin{align}
d x^{b}/dt =(X_{H_{s}})^{b},\quad b=1,\dots,\dim M,\label{eom1}\\[3mm]
d x^{b}/d\tilde t =(X_{\tilde H_{s}})^{b},\quad b=1,\dots,\dim M. \label{eom2}%
\end{align}

According to \cite{hiet} we have a \emph{reciprocal transformation} (see
e.g.\ \cite{cr2,cr1,rj} for more details on such transformations) relating the times
$t$ and $\tilde t$:
\begin{equation}
\label{rt0}d\tilde t=-H_{s}^{(1)} dt.
\end{equation}

When does (\ref{rt0}) turn (\ref{eom1}) into (\ref{eom2})? From (\ref{rt0}) we
find that
\[
d/dt =-H_{s}^{(1)} d/d\tilde t,
\]
and taking into account (\ref{eom1}) and (\ref{eom2}) we see that our question
boils down to the following: when does the equality
\begin{equation}
\label{td0}X_{H_{s}}=-H_{s}^{(1)} X_{\tilde H_{s}}%
\end{equation}
hold?

We have $X_{H_{s}}=P d H_{s}=P dH_{s}^{(0)}+\alpha Pd H_{s}^{(1)}$ and
\[
X_{\tilde H_{s}}=Pd\tilde H_{s}\overset{(\ref{tr})}{=} -\displaystyle\frac
{1}{H_{s}^{(1)}}P d H_{s}^{(0)}+\displaystyle\frac{H_{s}^{(0)}-\tilde\alpha
}{\left(  H_{s}^{(1)}\right)  ^{2}}P d H_{s}^{(1)}.
\]
Plugging this into (\ref{td0}) and multiplying the resulting equation by
$H_{s}^{(1)}$, which is nonzero by assumption, we obtain the following
equation:
\begin{equation}
\label{td}(H_{s}^{(0)}+\alpha H_{s}^{(1)}-\tilde\alpha)P d H_{s}^{(1)}=0.
\end{equation}

Clearly, (\ref{td}) holds if and only if either $P d H_{s}^{(1)}=0$ or
$(H_{s}^{(0)}+\alpha H_{s}^{(1)}-\tilde\alpha)=0$. The first possibility
immediately yields the following result:

\begin{prop}
\label{eomp0} Under the assumptions of Proposition \ref{trp}, suppose that
$H_{s}^{(1)}$ is a Casimir function for $P$, i.e., $Pd H_{s}^{(1)}=0$. Then
the transformation (\ref{rt0}) sends the equations of motion (\ref{eom1}) for
$H_{s}$ into the equations of motion (\ref{eom2}) for $\tilde H_{s}$.
\end{prop}

The second possibility is slightly more involved and
will be of greater interest to us in the sequel:

\begin{prop}
\label{eomp} Under the assumptions of Proposition \ref{trp} the transformation
(\ref{rt0}) sends the equations of motion (\ref{eom1}) for $H_{s}$ restricted
onto the level surface $H_{s}=\tilde\alpha$ of $H_{s}$ into the equations of
motion (\ref{eom2}) for $\tilde H_{s}$ restricted onto the level surface
$\tilde H_{s}=\alpha$ of $\tilde H_{s}$.

\end{prop}

\begin{proof}
Indeed, if $H_{s}=\tilde\alpha$ then $H_{s}^{(0)}+\alpha H_{s}^{(1)}%
-\tilde\alpha=0$ by (\ref{df}). On the other hand, the condition $\tilde
H_{s}=\alpha$ is equivalent to $H_{s}=\tilde\alpha$ by
(\ref{tr}). Thus, if $\tilde H_{s}=\alpha$ or $H_{s}=\tilde\alpha$ then
(\ref{td}) holds, and the result follows. \looseness=-1
\end{proof}

Note that as the parameters $\alpha$ and $\tilde\alpha$ are arbitrary,
(\ref{rt0}) will transform the equations of motion (\ref{eom1}) restricted
onto \emph{any} given level surface of $H_{s}$ into the equations of motion
(\ref{eom2}) restricted onto \emph{any} given level surface of $\tilde H_{s}$.
Thus we have a remarkable duality among the deformation parameters and the
energy (eigen)values that can be readily transferred to the quantum case.




\section{Multitime extension}

Now suppose that all $H_{i}$ are in involution:
\begin{equation}
\label{invol}\{H_{i},H_{j}\}=0,\quad i,j=1,\dots,k.
\end{equation}
Then by Corollary \ref{com} so are $\tilde H_{i}$:
\[
\{\tilde H_{i},\tilde H_{j}\}=0,\quad i,j=1,\dots,k.
\]
Consider the simultaneous evolutions
\begin{align}
d x^{b}/dt_{i} =(X_{H_{i}})^{b},\quad b=1,\dots,\dim M,\quad i=1,\dots
,k,\label{eomk1}\\[3mm]
d x^{b}/d\tilde t_{i} =(X_{\tilde H_{i}})^{b},\quad b=1,\dots,\dim M,\quad
i=1,\dots,k, \label{eomk2}%
\end{align}
and the following extension of (\ref{rt0}):
\begin{equation}
\label{rct}d\tilde t_{s}=-\sum\limits_{i=1}^{k} H_{i}^{(1)}d t_{i}%
,\qquad\tilde t_{i}=t_{i},\quad i=1,2,\dots,k,
\end{equation}
It is straightforward to verify that by virtue of (\ref{df}), (\ref{invol}),
and (\ref{eomk1}) we have
\[
\{H_{i}^{(0)},H_{j}^{(0)}\}=0,\quad\{H_{i}^{(1)},H_{j}^{(1)}\}=0, \quad
(H_{i}^{(1)})_{t_{j}}=(H_{j}^{(1)})_{t_{i}}, \quad i,j=1,\dots,k,
\]
so the transformation (\ref{rct}) for $\tilde t_{s}$ is well-defined. This is
where we need the commutativity of $H_{i}$.

By virtue of (\ref{rct}) we have
\[
\frac{d}{d t_{s}}=-H_{s}^{(1)}\frac{d}{d\tilde t_{s}},\qquad\frac{d}{d t_{i}%
}=\frac{d}{d\tilde t_{i}}-H_{i}^{(1)}\frac{d}{d\tilde t_{s}}, \quad
i=1,2,\dots,s-1,s+1,\dots,k.
\]
In view of (\ref{eomk1}) and (\ref{eomk2}) we search for conditions when
\begin{equation}
X_{H_{s}}=-H_{s}^{(1)} X_{\tilde H_{s}},\quad X_{H_{i}}=X_{\tilde H_{i}}%
-H_{i}^{(1)} X_{\tilde H_{s}}, \quad i=1,2,\dots,s-1,s+1,\dots,k .
\label{eeom1}%
\end{equation}

Plugging (\ref{df}) and (\ref{tr}) into (\ref{eeom1}) yields, after some
simplifications,
\[
\displaystyle\left(  H_{s}^{(0)}+\alpha H_{s}^{(1)}-\tilde\alpha\right)  P d
H_{i}^{(1)}=0, \quad i=1,\dots,k.
\]

Hence we have the following generalizations of Propositions~\ref{eomp0} and
\ref{eomp}:

\begin{prop}
\label{cas} Under the assumptions of Proposition \ref{trp}, suppose that
\[
\{H_{i},H_{j}\}=0,\quad i,j=1,\dots,k,
\]
and that $H_{i}^{(1)}$, $i=1,\dots,k$, are Casimir functions for $P$, i.e.,
$Pd H_{i}^{(1)}=0$, $i=1,\dots,k$. Then
(\ref{rct}) transforms (\ref{eomk1}) into (\ref{eomk2}).
\end{prop}


\begin{prop}
\label{eomp1} Under the assumptions of Proposition \ref{trp}, suppose that
\[
\{H_{i},H_{j}\}=0,\quad i,j=1,\dots,k.
\]
Then the reciprocal transformation (\ref{rct})
sends the equations of motion (\ref{eomk1}) restricted onto the level surface
$H_{s}=\tilde\alpha$
into the equations of motion (\ref{eomk2}) restricted onto the level surface
$\tilde H_{s}=\alpha$.

\end{prop}



\section{Canonical Poisson structure}

Let $P$ be a canonical Poisson structure on
$M=\mathbb{R}^{2n}$.
Then the Hamilton--Jacobi equations for $H_{i}$ and $\tilde H_{i}$ have a
common solution, cf.\ \cite{bkm}. Namely, we have the following extension of
the results of \cite{bkm} to the Hamiltonians that are not necessarily
quadratic in the momenta:

\begin{prop}
\label{shjetr} Under the assumptions of Proposition~\ref{trp}
let $M=\mathbb{R}^{2n}$, $P$ be a canonical Poisson structure on $M$, and
$\lambda_{i}$, $\mu_{i}$, $i=1,\dots,n$, be the Darboux coordinates for $P$,
i.e.,
$\{\lambda_{i},\mu_{j}\}=\delta_{ij}$.
Let $\boldsymbol{\lambda}=(\lambda_{1},\dots,\lambda_{n})$ and
$\boldsymbol{\mu}=(\mu_{1},\dots,\mu_{n})$.

Let $S=S(\boldsymbol{\lambda},\alpha, E_{s}, a_{1},\dots,a_{n-1})$, where
$a_{i}$ are arbitrary constants, be a complete integral of the stationary
Hamilton--Jacobi equation for the Hamiltonian $H_{s}=H_{s}(\alpha
,\boldsymbol{\lambda}, \boldsymbol{\mu})$,
\[
H_{s}(\alpha,\boldsymbol{\lambda}, \partial S/\partial\boldsymbol{\lambda
})=E_{s}.
\]

If we set $E_{s}=\tilde\alpha$ and $\alpha=\tilde E_{s}$ then
$S=S(\boldsymbol{\lambda},\alpha, E_{s}, a_{1},\dots,a_{n-1})$ is a complete
integral of the stationary Hamilton--Jacobi equation for the Hamiltonian
$\tilde H_{s}=\tilde H_{s}(\tilde{\alpha},\boldsymbol{\lambda},
\boldsymbol{\mu})$,
\[
\tilde H_{s}(\tilde{\alpha},\boldsymbol{\lambda}, \partial S/\partial
\boldsymbol{\lambda})=\tilde E_{s}.
\]

Moreover, let
$\{H_{i},H_{j}\}=0$, $i,j=1,\dots,k$, and let
\begin{equation}
\label{ci0}S=S(\boldsymbol{\lambda},\alpha, E_{1},\dots,E_{k}, a_{1}%
,\dots,a_{n-k})
\end{equation}
where $a_{i}$ are arbitrary constants, be a complete integral for the system
of stationary Hamilton--Jacobi equations
\[
H_{i}(\alpha,\boldsymbol{\lambda}, \partial S/\partial\boldsymbol{\lambda
})=E_{i}, \quad i=1,\dots,k.
\]
If we set
\[
\alpha=\tilde E_{s},\quad E_{s}=\tilde\alpha,\quad E_{i}=\tilde E_{i},\quad
i=1,2,\dots,s-1,s+1,\dots,k,
\]
then $S$ (\ref{ci0}) is also a complete integral for the system
\[
\tilde H_{i}(\tilde{\alpha},\boldsymbol{\lambda}, \partial S/\partial
\boldsymbol{\lambda})=\tilde E_{i}, \quad i=1,\dots,k.
\]

\end{prop}

As for the equations of motion, we have, in addition to general
Propositions~\ref{eomp} and \ref{eomp1}, a somewhat more explicit result:

\begin{cor}
\label{reoms} Under the assumptions of Proposition \ref{eomp1} let
$M=\mathbb{R}^{2n}$, $P$ be a canonical Poisson structure on $M$, and
$\lambda_{i}$, $\mu_{i}$, $i=1,\dots,n$ be the Darboux coordinates for $P$,
i.e.,
$\{\lambda_{i},\mu_{j}\}=\delta_{ij}$.
Let $\boldsymbol{\lambda}=(\lambda_{1},\dots,\lambda_{n})$ and
$\boldsymbol{\mu}=(\mu_{1},\dots,\mu_{n})$.

Suppose that $k=n$, $\partial H_{i}^{(1)}/\partial\boldsymbol{\mu}=0$ for all
$i=1,\dots,n$ and that $\lambda_{i}$ can be chosen as local coordinates on the
Lagrangian submanifold $N_{E}=\{ (\boldsymbol{\lambda}, \boldsymbol{\mu})\in
M| H_{i}(\alpha,\boldsymbol{\lambda}, \boldsymbol{\mu})=E_{i}, \quad
i=1,\dots,n\}$ (in other words, the system $H_{i}(\alpha,\boldsymbol{\lambda},
\boldsymbol{\mu})=E_{i}$, $i=1,\dots,n$ can be solved for $\boldsymbol{\mu}$),
and that we have
\begin{equation}
\label{par0}\tilde E_{s}=\alpha,\quad E_{s}=\tilde\alpha,\quad\tilde
E_{i}=E_{i},\quad i=1,2,\dots,s-1,s+1,\dots,n.
\end{equation}

Then the reciprocal transformation (\ref{rct}) turns the system
\begin{equation}
\label{lt1}d\boldsymbol{\lambda}/ dt_{i}=(\partial H_{i}/\partial
\boldsymbol{\mu})|_{N_{E}},\quad i=1,\dots,n,
\end{equation}
into
\begin{equation}
\label{lt2}d\boldsymbol{\lambda}/d\tilde t_{i} =(\partial\tilde H_{i}%
/\partial\boldsymbol{\mu})|_{\tilde N_{\tilde E}},\quad i=1,\dots,n,
\end{equation}
where $\tilde N_{\tilde E}=\{ (\boldsymbol{\lambda}, \boldsymbol{\mu})\in M|
\tilde H_{i}(\tilde{\alpha},\boldsymbol{\lambda}, \boldsymbol{\mu})=\tilde
E_{i}, \quad i=1,\dots,n\}$.
\end{cor}

For instance, if we have
\begin{equation}
\label{nat}H_{i}=\displaystyle\frac12(\boldsymbol{\mu}, G_{i}%
(\boldsymbol{\lambda})\boldsymbol{\mu}) +V_{i}(\boldsymbol{\lambda}) +\alpha
W_{i}(\boldsymbol{\lambda}), \quad i=1,\dots,n,
\end{equation}
where $(\cdot,\cdot)$ stands for the scalar product in $\mathbb{R}^{n}$ and
$G_{i} (\boldsymbol{\lambda})$ are $n\times n$ matrices, then the system
(\ref{lt1}) reads
\begin{equation}
\label{lm1}d\boldsymbol{\lambda}/ dt_{i}=G_{i}%
(\boldsymbol{\lambda })\boldsymbol{M},
\end{equation}
where $\boldsymbol{\mu}=\boldsymbol{M}(\boldsymbol{\lambda},\alpha,E_{1}%
,\dots,E_{n})$ is a general solution of the system $H_{i}(\alpha
,\boldsymbol{\lambda},\boldsymbol{\mu})=E_{i}$, $i=1,\dots,n$.

If we eliminate $\boldsymbol{M}$ from (\ref{lm1}) then we obtain the
\emph{dispersionless Killing systems} (cf.\ \cite{blasak})
\begin{equation}
\label{ki}\boldsymbol{\lambda}_{t_{i}}=G_{i} (G_{s})^{-1}%
\boldsymbol{\lambda }_{t_{s}},\quad i=1,2,\dots,s-1,s+1,\dots,n,
\end{equation}
and
the reciprocal transformation (\ref{rct}), which in our case reads
\[
d\tilde t_{s}=-\sum\limits_{i=1}^{n} W_{i}(\boldsymbol{\lambda})d t_{i}%
,\qquad\tilde t_{i}=t_{i},\quad i\neq s,
\]
turns (\ref{ki}) into
\begin{equation}
\label{ki2}\boldsymbol{\lambda}_{\tilde t_{i}}=\tilde G_{i} (\tilde
G_{s})^{-1}\boldsymbol{\lambda}_{\tilde t_{s}}, \quad i=1,2,\dots
,s-1,s+1,\dots,n,
\end{equation}
where $\tilde G_{s}=-G_{s}/W_{s}$ and $\tilde G_{i}=G_{i}-W_{i} G_{s}/W_{s}$,
$i=1,2,\dots,s-1,s+1,\dots,n$, are related to the Hamiltonians
\begin{equation}
\tilde H_{i}=\displaystyle\frac12(\boldsymbol{\mu},\tilde G_{i}%
(\boldsymbol{\lambda})\boldsymbol{\mu}) +\tilde V_{i}(\boldsymbol{\lambda})
+\tilde\alpha\tilde W_{i}(\boldsymbol{\lambda}), \quad i=1,\dots,n.
\end{equation}

\section{Solving the reduced equations of motion}

Under the assumptions of Corollary~\ref{reoms} we can apply
Proposition~\ref{shjetr} in order to obtain the solutions of equations of
motion (\ref{lt1}) and (\ref{lt2}) as follows:

\begin{cor}
\label{soleom} Under the assumptions of Corollary~\ref{reoms}, suppose that
\begin{equation}
\label{ci}S=S(\boldsymbol{\lambda},\alpha, E_{1},\dots,E_{n})
\end{equation}
is a complete integral for the system of stationary Hamilton--Jacobi
equations
\[
H_{i}(\alpha,\boldsymbol{\lambda}, \partial S/\partial\boldsymbol{\lambda
})=E_{i}, \quad i=1,\dots,n.
\]
Then a general solution of (\ref{lt1}) for $i=r$ can be written in implicit
form as
\begin{equation}
\label{is1}\partial S/\partial E_{j}=\delta_{jr} t_{r}+b_{j},\quad
j=1,\dots,n,
\end{equation}
where $b_{j}$ are arbitrary constants, and by virtue of (\ref{par0}) a general
solution of (\ref{lt2}) for $i=r$ can be written in implicit form as
\begin{equation}
\label{is2}\partial S/\partial\tilde E_{j}=\delta_{jr} \tilde t_{r}+ b_{j}
,\quad j=1,\dots,n.
\end{equation}

\end{cor}

Comparing (\ref{is1}) and (\ref{is2}) and using (\ref{par0}) we readily see
that, in perfect agreement with (\ref{rct}),
$t_{i}=\tilde t_{i}$ for $i\neq s$, but $t_{s}=\partial S/\partial
E_{s}- b_{s}=\partial S/\partial\tilde{\alpha}-b_s$ while $\tilde t_{s}=\partial
S/\partial\tilde E_{s}-b_s=\partial S/\partial\tilde{\alpha}-b_s$, so this approach
does not yield an \emph{explicit} formula expressing $\tilde t_{s}$ as a
function of $\boldsymbol{\lambda}, \boldsymbol{\mu}$, and $t_{i}$.


In order to find a complete integral (\ref{ci})
we can use separation of variables as follows (see e.g.\ \cite{mac2005} and
references therein). Under the assumptions of Corollary~\ref{soleom} suppose
that $\lambda_{i}$, $\mu_{i}$, $i=1,\dots,n$ are separation coordinates for
the Hamiltonians $H_{i}$, $i=1,\dots,n$, that is, the system of equations
$H_{i}(\alpha,\boldsymbol{\lambda}, \boldsymbol{\mu})=E_{i}$, $i=1,\dots,n$,
is equivalent to the following one:
\begin{equation}
\label{sr}\varphi_{i}(\lambda_{i},\mu_{i},\alpha, E_{1},\dots,E_{n})=0, \quad
i=1,\dots,n,
\end{equation}
that is, the separation relations on the Lagrangian submanifold $N_{E}$.

Consider the system of stationary Hamilton--Jacobi equations for $H_{i}$
\begin{equation}
\label{shje}H_{i}(\alpha,\boldsymbol{\lambda}, \partial S/\partial
\boldsymbol{\lambda})=E_{i}, \quad i=1,\dots,n.
\end{equation}
By the above, (\ref{shje}) is equivalent to the system
\begin{equation}
\label{sshje}\varphi_{i}(\lambda_{i},\partial S/\partial\lambda_{i},\alpha,
E_{1},\dots,E_{n})=0, \quad i=1,\dots,n.
\end{equation}
Suppose that (\ref{sr}) can be solved for $\mu_{i}$, $i=1,\dots,n$:
\[
\mu_{i}=M_{i}(\lambda_{i},\alpha,E_{1},\dots,E_{n}), \quad i=1,\dots,n.
\]

Then there exists a complete integral of (\ref{sshje}), and hence of
(\ref{shje}), of the form (cf.\ e.g.\ \cite{mac2005} and references therein)
\begin{equation}
\label{cis}S=\sum\limits_{i=1}^{n} \int M_{i}(\lambda_{i},\alpha,E_{1}%
,\dots,E_{n})d\lambda_{i},
\end{equation}
and general solutions for (\ref{lt1}) and (\ref{lt2}) can be found using the
method of Corollary~\ref{soleom}.

The formulas (\ref{is1}) take the form
\[
\sum\limits_{i=1}^{n} \int(\partial M_{i}(\lambda_{i},\alpha,E_{1},\dots
,E_{n})/\partial E_{j})d\lambda_{i} =\delta_{jr} t_{r}+b_{j},\quad
j=1,\dots,n.
\]

For $r=s$ we have
\[
\tilde t_{s}+b_s=\partial S/\partial\tilde E_{s}=\partial S/\partial\alpha
=\sum\limits_{i=1}^{n} \int(\partial M_{i}(\lambda_{i},\alpha,E_{1}%
,\dots,E_{n})/\partial\alpha)d\lambda_{i}.
\]


\section{Deformations of seed systems}

Under the assumptions of Corollary 2, suppose that $\lambda_{i}$, $\mu_{i}$,
$i=1,\dots,n$ are \emph{separation coordinates} for the Hamiltonians $H_{i}$,
$i=1,\dots,n$, then the Lagrangian submanifold $N_{E}$ is defined by $n$
separation relations (\ref{sr}). Further assume that all functions
$\varphi_{i}$
are identical,
\begin{equation}
\label{sk}\varphi_{i}=\varphi(\lambda_{i},\mu_{i},\alpha, E_{1},\dots,E_{n}),
\quad i=1,\dots,n.
\end{equation}
Then the relations (\ref{sr}) mean that the points $\lambda_{i},\mu_{i}$
belong to the separation curve
\begin{equation}
\label{sk1}\varphi(\lambda,\mu,\alpha, E_{1},\dots,E_{n})=0
\end{equation}
for all $i=1,\dots,n$.

If the relations
\[
\varphi(\lambda_{i},\mu_{i},\alpha, H_{1},\dots,H_{n})=0, \quad i=1,\dots,n,
\]
uniquely determine the Hamiltonians $H_{i}$ for $i=1,\dots,n$, then for the
sake of brevity we shall say that $H_{i}$ for $i=1,\dots,n$ have the
separation curve
\[
\varphi(\lambda,\mu,\alpha, H_{1},\dots,H_{n})=0.
\]
Fixing values of all Hamiltonians $H_{i}=E_{i}$, $i=1,\dots,n$,
picks a particular Lagrangian submanifold from the Lagrangian foliation.
Setting $\alpha=0$ in the above formulas we see that $\lambda_{i}$, $\mu_{i}$,
$i=1,\dots,n$ are separation coordinates for the Hamiltonians $H_{i}^{(0)}$,
$i=1,\dots,n$, as well,
so the system of equations $H_{i}^{(0)}(\boldsymbol{\lambda},
\boldsymbol{\mu
})=E_{i}$, $i=1,\dots,n$, is equivalent to
\begin{equation}
\label{srk}\varphi_{0}(\lambda_{i},\mu_{i},E_{1},\dots,E_{n})=0, \quad
i=1,\dots,n,
\end{equation}
where $\varphi_{0}(\lambda_{i},\mu_{i}, E_{1},\dots,E_{n}) =\varphi
(\lambda_{i},\mu_{i},\alpha, E_{1},\dots,E_{n})|_{\alpha=0}$.

In what follows we shall restrict ourselves
to the separable systems
whose separation curves $\varphi_{0}=0$ read
\begin{equation}
\label{ksk}H_{1}^{(0)}\lambda^{\beta_{1}}+H_{2}^{(0)}\lambda^{\beta_{2}%
}+\cdots+H_{n}^{(0)}\lambda^{\beta_{n}}=\psi(\lambda,\mu)
\end{equation}
where $n+k-1=\beta_{1}>\beta_{2}>...>\beta_{n}=0, k\in\mathbb{N}$ and
$\psi(\lambda,\mu)$ is a smooth function.
Each class of systems (\ref{ksk}) is labeled by a sequence $(\beta
_{1},...,\beta_{n})$ while a particular system from a class is given by a
particular choice of $\psi(\lambda,\mu)$. In particular, the choice
$\psi(\lambda,\mu)=\frac{1}{2}f(\lambda)\mu^{2}+\gamma(\lambda)$ yields the
well-known classical St\"ackel systems.

For $k=0$ there is only one class given by
\begin{equation}
\label{bsk}H_{1}^{(0)}\lambda^{n-1}+H_{2}^{(0)}\lambda^{n-2}+\cdots
+H_{n}^{(0)}=\psi(\lambda,\mu)
\end{equation}
which is precisely the Benenti class of St\"ackel systems \cite{ben93,ben97}
if $\psi(\lambda,\mu)=\frac{1}{2}f(\lambda)\mu^{2}+\gamma(\lambda)$. All these
systems separate in the same set of coordinates $(\lambda_{i},\mu_{i})$ by
construction. We shall refer below to the systems with the separation curve
(\ref{bsk}) 
as to the systems from the {\em seed class}.

In \cite{mac2005}
it was shown that
an arbitrary class of the 
systems
with the separation curve (\ref{ksk}) is obtained via the so-called $k$-hole
deformation
from the seed class (\ref{bsk}).

Below we demonstrate that an arbitrary $k$-hole deformation
is nothing but a sequence of $k$ St\"ackel transforms (\ref{tr}), and hence
all separable classes (\ref{ksk}) are St\"ackel-equivalent to the seed class
in the sense of \cite{bkm}. In order to do this we first introduce an
alternative notation for different classes (\ref{ksk}), which is more
convenient for further considerations as well as for the bi-Hamiltonian extension.

We shall call a polynomial of the form
$\sum\limits_{j=1}^{s} a_{m+j}\lambda^{n+k-(m+j)}$
a \emph{sub-chain of length $s$} if $a_{k}\neq0$ for $k=m+1,...,m+s$, and a
\emph{string of holes of length $s$} if $a_{k}=0$ for $k=m+1,...,m+s$.

Now consider a polynomial $\sum\limits_{i=1}^{n+k}a_{i}\lambda^{n+k-i}$, where
precisely $k$ of the coefficients $a_{i}$ equal zero, but $a_{1}\neq0$ and
$a_{n+k}\neq0$.
Denote $n$ non-vanishing coefficients $a_{i}$
by $(H_{1}^{(0)},...,H_{n}^{(0)})$, where $0\neq a_{1}=H_{1}^{(0)}$ and $0\neq
a_{n+k}=H_{n}^{(0)}$ by assumption. In what follows we also assume that if
$0\neq a_{k}=H_{k^{\prime}}^{(0)}$ and $0\neq a_{l}=H_{l^{\prime}}^{(0)}$ and
$k>l$ then $k^{\prime}>l^{\prime}$.

It is immediate that any class of separation curves (\ref{ksk}) is uniquely
determined by a sequence of sub-chains and strings of holes $(n_{1}%
,m_{1},n_{2},m_{2},...,m_{l-1},n_{l}), \sum\limits_{i=1}^{l}n_{i}=n,
\sum\limits_{i=1}^{l-1}m_{i}=k$, where $n_{i}$ is the length of $i$-th
sub-chain and $m_{i}$ is the length of $i$-th string of holes. The separation
curve corresponding to a sequence $(n_{1},m_{1},n_{2},m_{2},...,m_{l-1}%
,n_{l}), \sum\limits_{i=1}^{l}n_{i}=n, \sum\limits_{i=1}^{l-1}m_{i}=k$, reads
\begin{equation}
\label{kskn}\sum\limits_{j=1}^{n_{l}} H_{j}^{(0)}\lambda^{n+k-j}
+\sum\limits_{r=1}^{l-2}\sum\limits_{j=1}^{n_{l-r}} H_{j+\sum\limits_{s=1}^{r}
n_{l+1-s}}^{(0)}\lambda^{n+k -\sum\limits_{s=1}^{r} (n_{l+1-s}+m_{l-s})}
+\sum\limits_{j=1}^{n_{1}} H_{n-n_{1}+j}^{(0)}\lambda^{n_{1}-j}= \psi
(\lambda,\mu).
\end{equation}

For $k=0$ we have only one chain (the seed class). For $k=1$ there are $(n-1)$
different classes consisting of two sub-chains $(n_{1},1,n_{2})$, $n_{1}%
+n_{2}=n$, separated by one hole. For $k=2$ we have $\frac{1}{2}(n-1)(n-2)$
different classes, where $(n-1)$ of these classes
consist of two sub-chains separated by a
two-hole string $(n_{1},2,n_{2})$, $n_{1}+n_{2}=n$, while the remaining cases
consist of three sub-chains $(n_{1},1,n_{2},1,n_{3})$, $n_{1}+n_{2}+n_{3}=n$,
separated by single holes, and so on.

Now define the transformation from the $k$-hole case to the $(k+1)$-hole one.
Without loss of generality we can restrict ourselves to considering the
following subcases only:

\begin{itemize}
\item[(i)] $(n_{1},m_{1},n_{2},m_{2},...,m_{l-1},n^{\prime}_{l})
\rightarrow(n_{1},m_{1},n_{2},m_{2},...,m_{l-1}+1,n_{l})$,\quad$n_{l}%
=n^{\prime}_{l}$,

\item[(ii)] $(n_{1},m_{1},n_{2},m_{2},...,m_{l-1},n_{l}^{\prime})
\rightarrow(n_{1},m_{1},n_{2},m_{2},...,m_{l-1},n_{l},1,n_{l+1}),\quad
n_{l}+n_{l+1}=n_{l}^{\prime}.$
\end{itemize}

In the case (i) the number of sub-chains and their lengths are preserved,
while the length of the last string of holes is increased by one. In the case
(ii) the last sub-chain is split into two sub-chains by inserting an
additional hole.
%
Notice that
we can reach an arbitrary $s$-hole deformation in a unique way from the seed
class $(n)$ by applying the above recursion step $s$ times.

Passing from the $k$-hole deformation to the $(k+1)$-hole one means, according
to our recursion, that
for the separation curve we have
\[
H_{i}^{(0)}\lambda^{n+k-i} \rightarrow\bar H_{i}^{(0)}\lambda^{n+k-i+1}, \quad
i=1,\dots,n_{l+1}
\]
\[
H_{i}^{(0)}\lambda^{\beta_{i}} \rightarrow\bar H_{i}^{(0)}\lambda^{\beta_{i}},
\quad i>n_{l+1}.
\]
If $n_{l}=0$ and $n_{l+1}=n_{l}^{\prime}$ then we have the case (i) while for
$n_{l+1}<n_{l}^{\prime}$ we have the case (ii).

For the sake of convenience we now formally merge the cases (i) and (ii) into
a single transformation
\[
(n_{1},m_{1},n_{2},m_{2},\dots,m_{l-1},n'_{l}) \rightarrow(n_{1}%
,m_{1},n_{2},m_{2},\dots,m_{l-1},n_{l},1,n_{l+1}),\quad
n_{l}+n_{l+1}=n'_{l},
\]
where $n_{l}=0$ for the case (i) and
$1\leq
n_{l}\leq n_{l}^{\prime}-1$ for the case (ii). Here $n_{l}=0$
corresponds to a void sub-chain (sub-chain of zero length). \looseness=-1

\begin{prop}
Consider two $n$-tuples of Hamiltonians $\{H_{i}^{(0)}\}$ and $\{\bar{H}%
_{i}^{(0)}\}$ with the separation curves of the form
\begin{equation}
\sum\limits_{j=1}^{n_{l}^{\prime}}H_{j}^{(0)}\lambda^{n+k-j}+
\sum\limits_{r=1}^{l-2}\sum\limits_{j=1}^{n_{l-r}} H_{j+n'_l+\sum\limits_{s=2}^{r}
n_{l+1-s}}^{(0)}\lambda^{n+k -n'_{l}-m_{l-1}-\sum\limits_{s=2}^{r} (n_{l+1-s}+m_{l-s})}
+\sum
\limits_{j=1}^{n_{1}}H_{n-n_{1}+j}^{(0)}\lambda^{n_{1}-j}=\psi(\lambda
,\mu)\label{kskna}%
\end{equation}
corresponding to the $k$-hole deformation of the seed class (\ref{bsk}), and
\begin{equation}
\ba{l}
\sum\limits_{j=1}^{n_{l+1}}\bar{H}_{j}^{(0)}\lambda^{n+k+1-j}+\sum
\limits_{j=1}^{n_{l}}\bar{H}_{n_{l+1}+j}^{(0)}\lambda^{n+k-n_{l+1}-j}%
+
\sum\limits_{r=2}^{l-1}\sum\limits_{j=1}^{n_{l+1-r}} \bar H_{j+\sum\limits_{s=1}^{r}
n_{l+2-s}}^{(0)}\lambda^{n+k+1-\sum\limits_{s=1}^{r} (n_{l+1-s}+m_{l-s})}\\
+\sum\limits_{j=1}^{n_{1}}\bar{H}_{n-n_{1}+j}^{(0)}\lambda^{n_{1}-j}%
=\psi(\lambda,\mu),\label{kskn1}%
\ea
\end{equation}
corresponding to the $(k+1)$-hole deformation of the seed class (\ref{bsk}), respectively.
Here $n_{l}=0$ and $n_{l+1}=n'_{l}$ for the case (i) and $1\leq
n_{l}\leq n_{l}^{\prime}-1$ and $n'_{l}=n_{l+1}+n_{l}$ for the case
(ii); $m_l=1$ in both cases.
\looseness=-1

Then the St\"ackel transform from $\{H_{i}^{(0)}\}$ to $\{\bar H_{i}^{(0)}\}$
reads
\begin{align}
\bar H_{1}^{(0)}  &  =-\frac{1}{V_{n_{l+1}}^{(n+k)}}H_{n_{l+1}}^{(0)}%
,\nonumber\\
\bar H_{i}^{(0)}  &  =H_{i-1}^{(0)}-\frac{V_{i-1}^{(n+k)}}{V_{n_{l+1}}%
^{(n+k)}}H_{n_{l+1}}^{(0)},\quad i=2,\dots,n_{l+1},\label{7}\\
\bar H_{i}^{(0)}  &  =H_{i}^{(0)}-\frac{V_{i}^{(n+k)}}{V_{n_{l+1}}^{(n+k)}%
}H_{n_{l+1}}^{(0)},\quad i>n_{l+1},\nonumber
\end{align}
where $V_{i}^{(n+k)}$ are separable potentials
defined by the relation \cite{mac2005}
\begin{equation}\label{vk}
\lambda^{n+k}+\sum\limits_{j=1}^{n'_{l}}V_{j}^{(n+k)}\lambda^{n+k-j}+
\sum\limits_{r=1}^{l-2}\sum\limits_{j=1}^{n_{l-r}} V_{j+n'_l+\sum\limits_{s=2}^{r}
n_{l+1-s}}^{(n+k)}\lambda^{n+k -n'_{l}-m_{l-1}-\sum\limits_{s=2}^{r} (n_{l+1-s}+m_{l-s})}
+\sum\limits_{j=1}^{n_{1}}V_{n-n_{1}+j}^{(n+k)}\lambda^{n_{1}-j}=0
\end{equation}
that must hold for $\lambda=\lambda_{i}, i=1,\dots,n$.
\end{prop}

\begin{proof}
In order to compare the Hamiltonians we should reduce the separation curve
(\ref{kskn1}) for ${\bar H_{i}^{(0)}}$ to that for ${H_{i}^{(0)}}$.
To this end we
get rid of the highest monomial ${\bar H_{1}^{(0)}}\lambda^{n+k}$ in
(\ref{kskn1}) by expressing $\lambda^{n+k}$ from
(\ref{vk}).
Then comparing coefficients of the separation curves for ${\bar H_{i}^{(0)}}$
and ${H_{i}^{(0)}}$ yields
\begin{align*}
H_{i}^{(0)}  &  =\bar H_{i+1}^{(0)}-V_{i}^{(n+k)}\bar
H_{1}^{(0)},\quad i=1,\dots,n_{l+1}-1,\\H_{n_{l+1 }}^{(0)}  &
=-V_{n_{l+1}}^{(n+k)}\bar H_{1}^{(0)},\\H_{i}^{(0)}  &  = \bar
H_{i}^{(0)}-V_{i}^{(n+k)}\bar H_{1}^{(0)},\quad i=n_{l+1}+1,\dots
,n
\end{align*}
and (\ref{7}) follows.
\end{proof}

Notice that after the renumeration of Hamiltonians $\bar H_{i}^{(0)}$,
\begin{align}
\bar H_{1}^{(0)}  &  =\tilde H_{n_{l+1}}^{(0)},\nonumber\\
\bar H_{i}^{(0)}  &  =\tilde H_{i-1}^{(0)},\quad i=2,\dots,n_{l+1},\label{8}\\
\bar H_{i}^{(0)}  &  =\tilde H_{i}^{(0)},\quad i>n_{l+1},\nonumber
\end{align}
we deal with a particular case of Proposition~\ref{trp}, where $H_{i}%
^{(1)}=V_{i}^{(n+k)}$, $\tilde{\alpha}=0$, $s=n_{l+1}$. Thus, the reciprocal
transformation
\begin{equation}
d\tilde t_{n_{l+1}}=-\sum\limits_{j=1}^{n} V_{j}^{(n+k)}d
t_{j},\qquad\tilde
t_{i}=t_{i},\quad i\neq n_{l+1},\label{reci}%
\end{equation}
transforms the equations of motion for Hamiltonians ${H_{i}=H_{i}^{(0)}+\alpha
H_{i}^{(1)}}$ of the $k$-hole deformation restricted onto the level surface
$H_{n_{l+1}}=0$ into the equations of motion for Hamiltonians ${\tilde
H_{i}^{(0)}}$ of the $(k+1)$-hole deformation restricted onto the level
surface $\tilde H_{n_{l+1}}^{(0)}=\bar H_{1}^{(0)}=\alpha$.

We can readily extend $\tilde H_{i}^{(0)}$ to $\tilde H_{i}$ using (\ref{tr})
and the results of \cite{mac2005}.
Namely, let
\begin{align}
\tilde H_{i}^{(1)}  &  =\frac{V_{i}^{(n+k)}}{V_{n_{l+1}}^{(n+k)}}=\tilde
V_{i}^{(n+k-n_{l+1})}, \quad i\neq n_{l+1},\nonumber\\
\tilde H_{n_{l+1}}^{(1)}  &  =\frac{1}{V_{n_{l+1}}^{(n+k)}}=\tilde V_{n_{l+1}%
}^{(n+k-n_{l+1})}.\nonumber
\end{align}
Then the separation curves for ${H_{i}}$ and ${\tilde H_{i}}$ are of the form
\[
\alpha\lambda^{n+k}+H_{n_{l+1}}\lambda^{n+k-n_{l+1}}+\sum\limits_{i\neq
n_{l+1}}H_{i}\lambda^{\beta_{i}} =\psi(\lambda,\mu),
\]
\[
\tilde H_{n_{l+1}} \lambda^{n+k}+\tilde{\alpha}\lambda^{n+k-n_{l+1}}%
+\sum\limits_{i\neq n_{l+1}}\tilde
H_{i}\lambda^{\beta_{i}}=\psi(\lambda,\mu),
\]
and the reciprocal transformation (\ref{reci}) transforms the equations
of motion for Hamiltonians $H_{i}$ restricted onto the level surface
$H_{n_{l+1}}=\tilde\alpha$ into the equations of motion for Hamiltonians
$\tilde H_{i}$ restricted onto the level surface $\tilde H_{n_{l+1}}=\alpha$.

As the whole procedure is recursive, it means that the $n$-tuple
of integrable Hamiltonian dynamical systems described by the
separation curve (\ref{ksk}), restricted onto a given Lagrangian
submanifold, are related to an $n$-tuple of Hamiltonian dynamical
systems from the seed class (also restricted onto an appropriate
Lagrangian submanifold) via the sequence of reciprocal
transformations.

\section{Example}

As a simple illustration of the above results,
consider the H\'enon--Heiles system on a four-dimensional phase space with the
coordinates $(p_{1},p_{2},q_{1},q_{2})$ and canonical symplectic structure.
The corresponding Hamiltonian
\[
H_{1} =\frac{1}{2}p_{1}^{2}+\frac{1}{2}p_{2}^{2}+q_{1}^{3}+\frac{1}{2}%
q_{1}q_{2}^{2}-\alpha q_{1},
\]
is in involution with
\[
H_{2} =\frac{1}{2}q^{2}p_{1}p_{2}-\frac{1}{2}q^{1}p_{2}^{2}+\frac{1}{16}%
q_{2}^{4}+\frac{1}{4}q_{1}^{2}q_{2}^{2}-\frac{1}{4}\alpha q_{2}^{2}.
\]
Consider the following one-hole deformation (St\"ackel transform) of $H_{1}$ and $H_{2}$:
\begin{align*}
\widetilde{H}_{1}  &  =\frac{1}{2}\frac{1}{q_{1}}p_{1}^{2}+\frac{1}{2}\frac
{1}{q_{1}}p_{2}^{2}+q_{1}^{2}+\frac{1}{2}q_{2}^{2}-\widetilde{\alpha}\frac
{1}{q^{1}}\\[3mm]
\widetilde{H}_{2}  &  =\frac{1}{2}q_{2}p_{1}p_{2}-\frac{1}{2}q_{1}p_{2}%
^{2}-\frac{1}{8}\frac{q_{2}^{2}}{q_{1}}p_{1}^{2}-\frac{1}{8}\frac{q_{2}^{2}%
}{q^{1}}p_{2}^{2}-\frac{1}{16}q_{2}^{4}+\widetilde{\alpha}\frac{1}{4}%
\frac{q_{2}^{2}}{q^{1}}.
\end{align*}

The corresponding reciprocal transformation
\[
d\widetilde{t_{1}}=q_{1}dt_{1}+\frac{1}{4}q_{2}^{2}dt_{2},\ \ \ \widetilde
{t_{2}}=t_{2}
\]
defines the map between equations of motion restricted to the respective
Lagrangian submanifolds, $N_{E}=\{H_{1}=\widetilde{\alpha},H_{2}=E\}$ and
$N_{\tilde E}=\{\widetilde{H}_{1}=\alpha,\widetilde{H}_{2}=E\}$.
%

The separation coordinates $(\lambda,\mu)$ are related to $p$'s and $q$'s by
the formulas
\[
\begin{array}
[c]{l}%
q_{1}=\lambda_{1}+\lambda_{2},\qquad q_{2}=2\sqrt{-\lambda_{1}\lambda_{2}%
},\\[5mm]%
\displaystyle p_{1}=\frac{\lambda_{1}\mu_{1}}{\lambda_{1}-\lambda_{2}}%
+\frac{\lambda_{2}\mu_{2}}{\lambda_{2}-\lambda_{1}},\qquad p_{2}%
=\sqrt{-\lambda_{1}\lambda_{2}}\left(  \frac{\mu_{1}}{\lambda_{1}-\lambda_{2}%
}+\frac{\mu_{2}}{\lambda_{2}-\lambda_{1}}\right) ,
\end{array}
\]
are common for $H_{1},H_{2}$ and $\tilde H_{1}$, $\tilde H_{2}$ and the
respective separation curves read
\begin{align*}
\alpha\lambda^{2}+\lambda H_{1}+H_{2}  &  =\frac{1}{2}\lambda\mu^{2}%
+\lambda^{4},\\[3mm]
\widetilde{H}_{1}\lambda^{2}+\widetilde{\alpha}\lambda+\widetilde{H}_{2}  &
=\frac{1}{2}\lambda\mu^{2}+\lambda^{4}.
\end{align*}



\section*{Acknowledgments}

This research was partially supported by the Czech Grant Agency (GA\v{C}R)
under grant No.\ 201/04/0538, by the Ministry of Education, Youth and Sports
of the Czech Republic (M\v{S}MT \v{C}R) under grant MSM 4781305904, and by the
Polish State Committee For Scientific Research (KBN) under the KBN Research
Grant No.\ 1 PO3B 111 27. M.B. is pleased to acknowledge kind hospitality of
the Mathematical Institute of Silesian University in Opava.
\looseness=-1


\end{document}